\documentclass{svproc}
\usepackage{url}

\usepackage{color}
\usepackage{amsmath,amssymb}
\usepackage{algorithm}
\usepackage{algpseudocode}
\usepackage{graphicx}
\usepackage{subfigure}

\begin{document}
\mainmatter              
%
\title{A framework for the analysis of supervised discrete event systems under attack}
\titlerunning{Supervised discrete event systems under attack}  
%
\author{Qi Zhang\inst{1} \and Carla Seatzu\inst{2} \and
Zhiwu Li\inst{1} \and Alessandro Giua\inst{2}}
\authorrunning{Qi Zhang et al.} 
%
\tocauthor{Ivar Ekeland, Roger Temam, Jeffrey Dean, David Grove,
Craig Chambers, Kim B. Bruce, and Elisa Bertino}
\institute{Xidian University, Xi'an 710071, China,\\
\email{qzhang\_3@stu.xidian.edu.cn, zhwli@xidian.edu.cn},
\and
University of Cagliari, 09123 Cagliari, Italy,\\
\email{\{seatzu, giua\}@diee.unica.it}}

\maketitle              

\begin{abstract}
This paper focuses on the problem of cyber attacks for discrete event systems under supervisory control. In more detail, the goal of the supervisor, who has a partial observation of the system evolution, is that of preventing the system from reaching a set of unsafe states. An attacker may act in two different ways: he can corrupt the observation of the supervisor editing the sensor readings, and can enable events that are disabled by the supervisor. This is done with the aim of leading the plant to an unsafe state, and keeping the supervisor unaware of that before the unsafe state is reached. A special automaton, called attack structure is constructed as the parallel composition of two special structures. Such an automaton can be used by the attacker to select appropriate actions (if any) to reach the above goal, or equivalently by the supervisor, to validate its robustness with respect to such attacks.
\end{abstract}

\section{INTRODUCTION}\label{Introduction}
Cyber-physical systems (CPS) operating in a feedback loop are particularly vulnerable to attacks since the communication between controllers and processes typically occur via a network such as the internet. Computers collect the data from the processes through sensors and based on such data provide a suitable control input \cite{Harirchi:2018}. Malicious attackers may corrupt the sensor readings collected by the controller, and/or may alter the control commands \cite{Clark:2017}. In this paper, we consider a plant $G$ modeled by a partially-observed discrete event system. The open-loop behaviour of the plant, i.e., the language $L(G)$, may contain undesirable strings that should be prevented by a feedback controller, thus a \emph{partial-observation supervisor} $S_P$ is introduced to restrict $L(G)$ within a sublanguage $K \subseteq L(G)$ by appropriately disabling events in the plant. As a special case, one may also consider the problem of avoiding a set of \emph{unsafe states} $X_{us} \subseteq X$, where $X$ is the state space of $G$.

In this paper we deal with the problem of cyber attacks in supervisory control systems. Two different kinds of attacks are considered: {\em sensor attack} and {\em actuator enablement attack}. In the first case the attacker may corrupt the sensor channels transmitting erroneous observations to the supervisor. In particular, the attacker may erase the output symbols produced by certain events, and may insert observations corresponding to events that have not occurred. In the second case the attacker corrupts the control commands of the supervisor enabling events that are disabled by the supervisor. This may lead the plant to an unsafe state and the supervisor may even not realize it, namely the attacker remains stealthy.

The problem of attack detection in CPS is attracting an increasing attention in the automatic control community. It has been investigated in \cite{Sundaram:2011}, \cite{Pasqualetti:2013}, \cite{Fawzi:2014} in the case of time-driven continuous systems. Interesting contributions have also been proposed in the discrete event systems framework. As an example, in \cite{Thorsley:2006} the authors determine the condition under which the supervisor can detect the presence of an attacker and prevent the system from generating illegal strings. They use a language measure technique to assess the damage caused by the attacker if the supervisor cannot block the intrusion, and determine an optimal specification for the supervisor to realize in the presence of an attacker. Other significant contributions on cyber attacks in supervisory control systems are \cite{Carvalho:2016}, \cite{Lima:2017}, \cite{Su:2017}, \cite{Goes:2017}, \cite{Lima:2018}, where the authors either consider sensor or actuator enablement attacks, and propose defensive strategies.

Our work has some similarities with \cite{Su:2018}, \cite{Wakaiki:2018}, \cite{Lin:2019}, \cite{Zhu:2019}, \cite{Carvalho:2018}, \cite{Lima:2019}, \cite{Qi}. However, in \cite{Su:2018}, \cite{Wakaiki:2018}, the authors only consider sensor attacks, and define a supervisor that is robust against such attacks \cite{Su:2018}. In \cite{Lin:2019}, \cite{Zhu:2019}, the authors only consider actuator enablement attacks, and develop a behavior-preserving supervisor that is robust against such attacks \cite{Zhu:2019}. In our work, we assume that sensor and actuator enablement attacks may occur simultaneously. The simultaneous occurrence of the two kinds of attacks has also been considered in \cite{Carvalho:2018}, \cite{Lima:2019}. However, in these papers, despite of us, authors do not guarantee that the attacker remains stealthy. This paper may be seen as an extension of our results in \cite{Qi} where we investigated the problem of cyber attacks at the observation layer, without supervisor, and stealthyness of the attacker is guaranteed.
We finally mention the very recent contributions in \cite{Carvalho:2018} and \cite{Lima:2019}. In particular, a diagnoser-based algorithm is developed in \cite{Carvalho:2018}, and a security module against cyber attacks is provided in \cite{Lima:2019}.

The solution proposed in this paper to derive a stealthy attack policy associated with both sensor and actuator enablement attacks, is based on the notion of \emph{attack structure} that simultaneously keep into account the set of states that are consistent with the real observation generated by the plant and the set of states that are consistent with the corrupted observation received by the supervisor. Such an attack structure is computed as the parallel composition of two particular structures, called attacker observer and supervisor under attack. This verifies the effectiveness of an attacker that is defined based on it. A way to refine it is provided, selecting the so called \emph{supremal stealthy attack substructure}. The attack structure may also be used from another perspective. Indeed, it allows the supervisor to analyze its robustness with respect to sensor and actuator enablement stealthy attacks.

We conclude this section pointing out that the problem considered in this paper belongs to an important class of problems in the framework of discrete event systems which can be addressed reconstructing the event sequence or the state trajectory in the presence of a partial observation of the system evolution and/or a partial knowledge of the system state. Other problems in this framework are: state estimation and control \cite{Giua:2014}, fault diagnosis \cite{Cabasino:2010,Cabasino:2011,Cabasino:2012}, and opacity analysis \cite{Tong:2017}. All such problems have been extensively investigated in the literature in the last decades and effective approaches have been developed for their solution. Some of such approaches can be appropriately adapted for the solution of problems of intrusion detection, and can be of inspiration for them. An example is the recent contribution by some of the authors of this paper \cite{Gao:2019} where it is shown how a particular intrusion detection problem can be converted into a problem of fault diagnosis, and consequently it can be solved using appropriate techniques in this framework. However, there are some key features of the intrusion detection problem that significantly distinguish it from the other problems. The first one is the fact that it involves two players that compete for different goals, the attacker and the plant observer/controller. On the contrary, in the other cases, only one actor is involved, namely, the state observer, the controller or the diagnoser. The only other exception in this respect occurs in the case of opacity where we also have an attacker who tries to discover some secrets. The second feature is that in all the problems mentioned above apart from intrusion detection, the observation mask is static, while in the case of intrusion detection, the observation mask is dynamically changing following a malicious strategy aiming to lead the system to a dangerous or undesirable condition.

The remainder of the paper is organized as follows. In Section~\ref{Preliminaries}, some necessary preliminaries on finite state automata and supervisory control theory are given. In Section~\ref{Attackmodel}, the attack model is introduced. In Section~\ref{Problemstatement}, the problem statement is presented. In Section~\ref{Plantsupervisorunderattack}, the attacker observer and the supervisor under attack are introduced. In Section~\ref{Sectionattackstructure}, we introduce the attack structure, which provides the basic tool for solving the problem formalized in Section~\ref{Problemstatement}. Indeed, it allows to select on-line an effective attacker. In Section~\ref{Sectionattackstructure}, it is also shown how to extract a supremal stealthy attack substructure starting from the attack structure. In Section~\ref{Conclusions}, conclusions are finally drawn and our future lines of research in this framework are pointed out.

\section{PRELIMINARIES}\label{Preliminaries}

A \emph{deterministic finite-state automaton} (DFA) is a 4-tuple $G=(X,E,\Delta,x_0)$, where \emph{X} is the finite set of states, \emph{E} is the alphabet of events, $\Delta \subseteq X \times E \times X$ is the transition relation, and $x_0$ is the initial state. The transition relation determines the dynamics of the DFA: if $(x,e,x') \in \Delta$, it means that the occurrence of event $e$ at state $x$ yields state $x'$. The transition relation can be extended to $\Delta^* \in X \times E^* \times X$: if $(x_i, \sigma, x_j) \in \Delta^*$ it means that  there exists a sequence of states $x_i,...,x_j \in X$ and a sequence of events  $\sigma=e_i...e_{j-1} \in E^*$ such that
$(x_i,e_i,x_{i+1}),...,(x_{j-1},e_{j-1},x_j) \in \Delta$. The \emph{language generated by} \emph{G} is defined as $L(G)=\{\sigma \in E^* | \exists x \in X: (x_0,\sigma,x) \in \Delta^*\}$. We denote as $\Gamma(x)=\{e \in E \ | \ \exists x' \in X: (x,e,x') \in \Delta\}$ the set of events that are \emph{active} at state $x$.

Given two alphabets $E' \subseteq E$, the \emph{natural projection} on $E'$, $P_{E'}:E^*\rightarrow E'$ is defined as \cite{Wonham:1989}:

\begin{equation} \label{enaturalprojection}
P_{E'}(\varepsilon):=\varepsilon \ \text{\ and \ } \
P_{E'}(\sigma e) := \left\{
             \begin{array}{lcl}
             {P_{E'}(\sigma) e} \ \text{if} \ e \in E', \\
             {P_{E'}(\sigma)} \ \text{if} \ e \in E \backslash E'.
             \end{array}
        \right.
\end{equation}

In simple words, given a string $\sigma \in E^*$, its natural projection on $E'$ is obtained by simply removing events that do not belong to $E'$. For simplicity, we use $P:E^*\rightarrow E_o^*$ to denote the natural projection on $E_o$.

Given two automata $G_1=(X_1,E_1,\Delta_1,x_{01})$ and $G_2=(X_2,E_2,\Delta_2,x_{02})$, the \emph{parallel composition} of $G_1$ and $G_2$ is denoted as $G=G_1 \| G_2=(X_1 \times X_2, E_1 \cup E_2, \Delta, (x_{01} \times x_{02}))$, where the transition relation $\Delta$ is defined as follows:

\begin{equation} \label{eparallelcomposition}
\left\{ \begin{array}{ll} \exists ((x_1,x_2),e,(x_1',x_2')) \in \Delta \text{ if } \exists (x_1,e,x_1') \in \Delta_1 \wedge  \exists (x_2,e,x_2') \in \Delta_2, \\  \exists ((x_1,x_2),e,(x_1',x_2)) \in \Delta \text{ if } \exists (x_1,e,x_1') \in \Delta_1 \wedge  e \notin E_2, \\ \exists ((x_1,x_2),e,(x_1,x_2')) \in \Delta \text{ if } \exists (x_2,e,x_2') \in \Delta_2 \wedge  e \notin E_1, \\ \text{undefined} \qquad \qquad \qquad \quad \ \ \text{otherwise}.  \end{array} \right.
\end{equation}

Consider a plant modeled by a DFA $G=(X,E,\Delta,x_0)$, let $E=E_o\cup E_{uo}=E_c\cup E_{uc}$, where $E_o$ is the set of \emph{observable events}, $E_{uo}$ is the set of \emph{unobservable events}, $E_c$ is the set of \emph{controllable events}, and $E_{uc}$ is the set of \emph{uncontrollable events}. The \emph{unobservable reach} of state $x$ is defined by a set of states $x' \in X$ reached from state $x \in X$ by executing an unobservable string $\sigma \in E_{uo}^*$, namely, $UR(x)=\{x' \in X \ | \ \exists \sigma \in E_{uo}^*: (x,\sigma,x') \in \Delta^*\}$.

Given a plant $G=(X,E,\Delta,x_0)$ with set of observable events $E_o$, the \emph{observer} of $G$ \cite{Lafortune:2008} is the DFA $Obs(G)=(B,E_o,\Delta_o,b_0)$, where:

\begin{itemize}
\item $B \subseteq 2^X$ is the set of states,

\item $\Delta_o:B \times E_o \rightarrow B$ is the transition relation defined as:

$\Delta_o(b,e_o):=\cup_{x\in b} UR(\{x'\ | \ \Delta(x,e_o)=x'\})$,

\item $b_0:=UR(x_0)$ is the initial state.
\end{itemize}

A \emph{partial-observation supervisor} $S_P=(Y,E,\Delta_s,y_0)$ can be finitely represented by a \emph{control function} $f_c$: $P[L(G)] \rightarrow 2^E$, where $P$ is the natural projection on $E_o$. Without loss of generality, we assume that $L(S_P) \subseteq L(G)$. Given a string $\sigma \in E^*$: $(x_0,\sigma,x) \in \Delta^*$, let $\xi=f_c[P(\sigma)]$ be the \emph{control input} by $S_P$ at state $x$. The resulting closed-loop system is an automaton denoted as $S_P/G=G \| S_P$, with generated language $L(S_P/G)$.

\section{THE ATTACKER: DEFINITIONS AND ASSUMPTIONS}\label{Attackmodel}

\begin{figure}[htbp]
  \centering
  \includegraphics[width=2.1in]{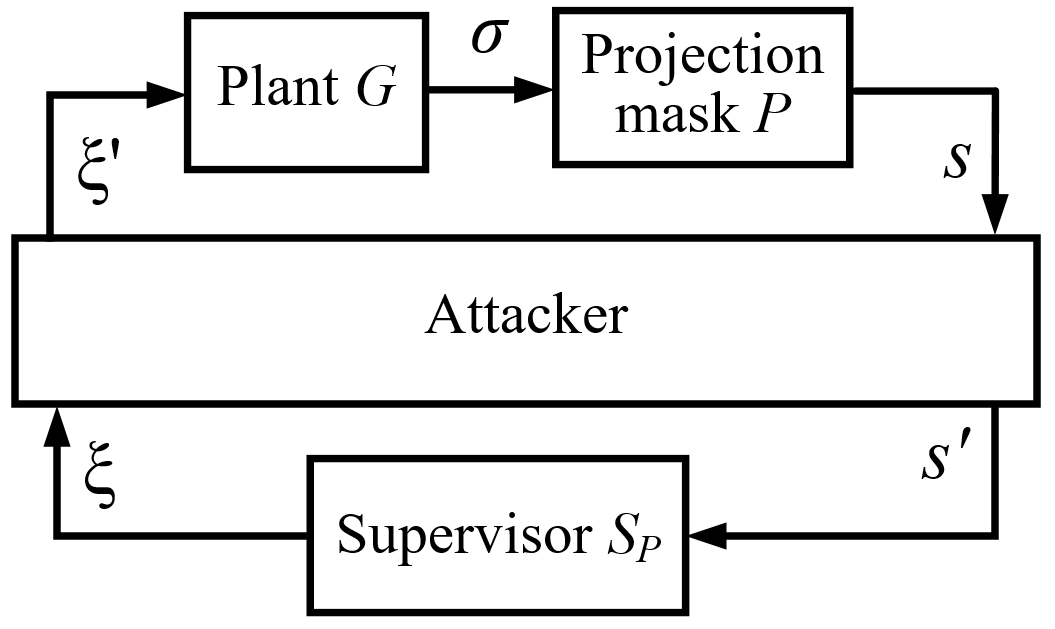}\\
  \centering
  \caption{Closed-loop system under attack.}
  \label{fig1}
\end{figure}

In this paper we consider a closed-loop system $S_P/G$ subject to attack according to the scheme in Fig.~\ref{fig1}. If $\sigma$ is a generic string generated by the plant, the \emph{observed string} is $s=P(\sigma)$. An attacker may corrupt the observation (sensor attack), inserting fake observations or erasing some output signals produced by events that have actually happened. Such a \emph{corrupted observation} is denoted by $s'$ and is still a sequence of events in $E_o$. The supervisor constructs its \emph{state estimation} based on $s'$ and elaborates its control input based on it. In addition, the attacker may also enable some events that are disabled by $S_P$ (actuator enablement attack). We denote as $\xi \in 2^E$ the \emph{control input} computed based on $s'$, and denote as $\xi' \in 2^E$ the \emph{corrupted control input} that actually restricts the behavior of $G$.

The set of \emph{compromised events}, namely the events that can be corrupted by the attacker, is denoted by $E_{com}$. Set $E_{com}$ includes: the events that the attacker may insert in the supervisor observation even if they have not actually occurred, the events that the attacker may erase in the supervisor observation, and the events that the attacker may enable even if they were disabled by the supervisor. The above three sets are denoted respectively, $E_{ins} $, $E_{era} $, and $E_{ena}$. In particular, it is $E_{ins} ,E_{era}  \subseteq E_o$ and $E_{ena} \subseteq E_c$. We assume that such sets are not necessarily disjoint.

We notice that the definition of compromised events has been firstly proposed in \cite{Goes:2017}. However, while in \cite{Goes:2017} the authors only consider sensor attacks, thus events in $E_{com}$ may only be of the first two types mentioned above, here we slightly generalize such a definition also dealing with actuator enablement attacks. As a result, events in $E_{com}$ may also be of the third type above.

Even if it is possible to define the sensor attack via a function that turns the observed string $s$ into the corrupted observation $s'$, for reasons that will be clear later, we prefer to formalize the problem in terms of a new string that is defined on an alphabet called \emph{attack alphabet}.

The \emph{attack alphabet} is defined as $E_a=E_o \cup E_+ \cup E_- \cup E_S$, and we assume that $E_o$, ${E_+}$, ${E_-}$, and $E_S$ are disjoint sets. The set of \emph{inserted events} is denoted as ${E_+}$, namely $E_+=\{e_+ \ | \ e \in E_{ins}\}$. The occurrence of $e_+ \in E_+$ implies that the attacker inserts in the supervisor observation an event $e$ that has not actually happened. The set of \emph{erased events} is denoted by ${E_-}$, namely $E_-=\{e_- \ | \ e \in E_{era}\}$. The occurrence of $e_- \in E_-$ means that the attacker erases from the supervisor observation event $e$ occurred in the plant. The set of \emph{enabled events} is denoted as $E_S$, namely $E_S=\{e_S \ | \ e \in E_{ena}\}$. The occurrence of $e_S \in E_S$ indicates that event $e$ disabled by the supervisor is again enabled by the attacker.

In addition we assume that the following relationship holds: $E_c \subseteq E_o$, namely all the events that are controllable are also observable. This implies that $E_{uo} \subseteq E_{uc}$, namely it could not happen that the attacker enables unobservable events, which simplifies our problem.

The following definition formalizes the notion of attacker via the sensor and actuator enablement attack functions.

\begin{definition} Consider a closed-loop system $S_P/G$ where $G=(X,E,\Delta,x_0)$, the set of compromised events is $E_{com}=E_{ins} \cup E_{era} \cup E_{ena}$, and the set of observable events is $E_o$. Let $Obs(G)=(B,E_o,\Delta_o,b_0)$ be the observer. An attacker is defined by a sensor attack function $f_{sen}: E_o^* \rightarrow E_a^\ast$, and an actuator enablement attack function $f_{ena}: \Xi \rightarrow \Xi'$, where $E_a$ is the attack alphabet, $\Xi \subseteq 2^E$ is the set of control inputs, and $\Xi' \subseteq 2^E$ is the set of corrupted control inputs. The attack functions satisfy the following conditions:

\begin{itemize}
\item [(a)] $f_{sen}(\varepsilon)\in E_+^*$;
\item [(b)] $\forall se \in E_o^*$ with $s \in E_o^*$:
\begin{equation}\label{equationb}
\left\{ \begin{array}{ll} f_{sen}(se) \in f_{sen}(s)\{e_-,e\}E_+^* \text { if } e \in E_{era}, \\  f_{sen}(se) \in f_{sen}(s)eE_+^* \text{ if } e \in E_o \setminus E_{era},  \end{array} \right.
\end{equation}
where $eE_+^* =\{ew_+ \ | \ w_+ \in E_+^*\}$ and $e_-E_+^* =\{e_-w_+ \ | \ w_+ \in E_+^*\}$;

\item [(c)] $\forall \xi \in 2^E: \xi'=f_{ena}(\xi) \subseteq \xi \cup E_{ena}$.~\hfill \qed

\end{itemize}
\label{def1}
\end{definition}

In Definition~\ref{def1}, condition (a) means that the attacker can insert any string in $E_+^*$ when no event has occurred in the plant. Condition (b) indicates that if an event $e \in E_{era}$ happens, the attacker can either erase event $e$ or not erase it, and then insert any string in $E_+^*$. If an event $e \!\in\! E_o \backslash E_{era}$ happens, the attacker can insert any string in $ E_+^*$ after $e$.
Condition (c) implies that the attacker can enable events in $E_{ena}$ that are not enabled by the supervisor.

The language modified by the attack functions is called {\em attack language} and is denoted by $L_a(S_P/G)$. We use $w$ to denote a string in $L_a(S_P/G)$, and we call $w$ \emph{attack string}.

\begin{definition} The supervisor projection $\hat{P}: E_a^* \rightarrow E_o^*$ is defined as:
\label{def2}
\end{definition}

\begin{equation}\label{esupervisorprojection}
\hat{P}(\varepsilon):=\varepsilon \ \text{\ and \ }
\hat{P}(w e_a):=\left\{
             \begin{array}{lcl}
             {\hat{P}(\!w\!)e} \ \text{if} \ e_a \!\in\! E_a \setminus E_-, \\
             {\hat{P}(\!w\!)} \ \text{if} \ e_a \!\in\! E_-.
             \end{array}
        \right.
\end{equation}~\hfill \qed

The projection describes how the supervisor deals with events in $E_a$, thus $\hat{P}$ is called supervisor projection. Namely, the supervisor cannot distinguish events in $E_+ \cup E_S$ from events in $E_o$, and the supervisor cannot observe events in $E_-$.

\begin{figure}[htbp]
  \centering
  \includegraphics[width=2.4in]{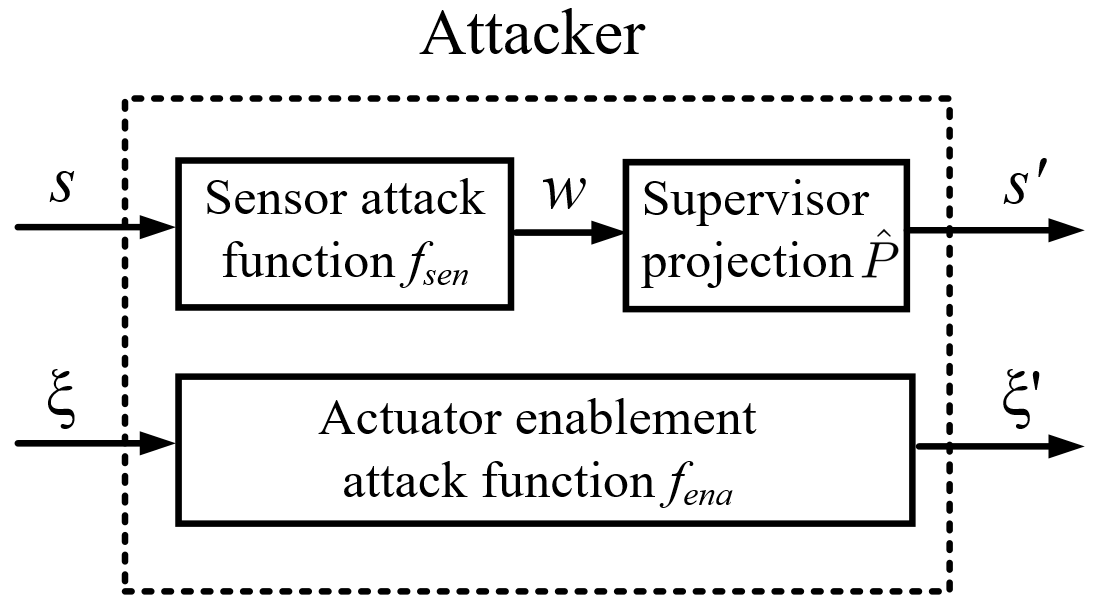}\\
  \caption{Internal structure of the attacker in Fig.~\ref{fig1}.}
  \label{fig2}
\end{figure}

The internal structure of the attacker that in Fig.~\ref{fig1} is represented as a black box having the observation $s$ and the control input $\xi$ as inputs, and the corrupted observation $s'$ and the corrupted control input $\xi'$ as outputs, is depicted in Fig.~\ref{fig2}.

Fig.~\ref{fig2} shows how the (uncorrupted) observation $s$ is corrupted by the sensor attack function $f_{sen}$, producing the attack string $w \in E_a^*$. Such a sequence is projected via the supervisor projection $\hat P$ on $E_o$, generating a string $s'$. The supervisor constructs its state estimation based on $s'$ and computes its control input $\xi$. Such a control input is altered by the actuator enablement attack function $f_{ena}$ producing the corrupted control input $\xi'$, which actually restricts the behavior of the plant.

\section{PROBLEM STATEMENT}\label{Problemstatement}

We first introduce some key definitions that will be useful in the following.

\begin{definition} Consider a closed-loop system $S_P/G$ under attack. Let $L(S_P/G)$ be the closed-loop language under no attack and $L_a(S_P/G)$ be the attack language. The attacker is \emph{stealthy} if $\hat{P}[L_a(S_P/G)] \subseteq P[L(S_p/G)]$.~\hfill \qed
\label{def3}
\end{definition}

In simple words, an attacker is stealthy if the set of words that a supervisor may observe when the system is under attack is contained in the set of words the supervisor may observe when no attack occurs.

\begin{definition} The attacker projection $\tilde{P}: E_a^* \rightarrow E_o^*$ is defined as:
\label{defplantprojection}
\end{definition}

\begin{equation}\label{eplantprojection}
\tilde{P}(\varepsilon):=\varepsilon, \ \text{\ and \ }
\tilde{P}(w e_a):=\left\{
             \begin{array}{lcl}
             {\tilde{P}(w)e} \ \text{if} \ e_a \in E_a \setminus E_+, \\
             {\tilde{P}(w)} \ \text{if} \ e_a \in E_+.
             \end{array}
        \right.
\end{equation}~\hfill \qed

The projection describes how the attacker deals with events in $E_a$, thus $\tilde{P}$ is called attacker projection, and the projection returns the real observation. Namely, the attacker knows that events $e_a \in E_S$ are events that actually occurs in the plant, and events $e_a \in E_-$ are events that have been erased, thus he updates his state the same way in case of $e \in E_o$, $e_- \in E_-$, and $e \in E_S$. Since the attacker knows that $e_a \in E_+$ are fake events that have not actually occurred in the plant, then he does not update his state when $e_a \in E_+$ occurs.

We assume that a set of unsafe states $X_{us}$ is given, which corresponds to an undesirable or dangerous condition for the plant $G$. The supervisor controls the plant with the objective of preventing it from reaching the unsafe state. The goal of the attacker is that of preventing the supervisor from reaching his objective. The following definition provides a criterion to assess the attacker.

\begin{definition} Consider a closed-loop system $S_P/G$ where $G=(X,E,\Delta,x_0)$. Let $X_{us}$ be a set of unsafe states and $L_a(S_P/G)$ be the attack language. Let $Obs(G)=(B,E_o,\Delta_o,b_0)$ be the observer where $E_o$ is the set of observable events. An attacker is:

\begin{itemize}
\item
effective if $\exists w \in L_a(S_P/G)$: $(b_0,\tilde{P}(w), b) \in \Delta_o^*$ and $b \cap X_{us}\neq\emptyset$;

\item
ineffective if it is not effective.~\hfill \qed
\end{itemize}

\label{def4}
\end{definition}

In simple words, an attacker is effective if there exists an attack string $w \in L_a(S_P/G)$ such that executing $\tilde{P}(w)$ starting from the initial state $b_0$, the observer reaches a state $b$ that contains an element in the set of unsafe states. This means that the plant may reach an unsafe state when the attack string $w$ is generated.

In this paper, given a closed-loop system $S_P/G$ with set of compromised events $E_{com}$, we want to provide a criterion to establish if an attack strategy exists, which leads the plant to the unsafe state, and the supervisor cannot detect the presence of an attacker before the unsafe state is reached. Dually, the strategy can be used to validate the safeness of the system against such attacks.

\section{ATTACKER OBSERVER AND SUPERVISOR UNDER ATTACK}\label{Plantsupervisorunderattack}

In this section we introduce two special structures, called \emph{attacker observer} and \emph{supervisor under attack}, which provide the basis for the solution of the problem formulated in the previous section.

\subsection{Attacker observer}\label{Plantunderattack}

The attacker observer, denoted as $Obs_{att}(G)$, provides the real state estimation of the plant based on the attack strings $w\in E_a^*$. The attacker observer $Obs_{att}(G)$ describes all possible attack strings and the corresponding sets of consistent states of the plant. Since attacks are performed by the attacker, he knows which observations originate from events that have really occurred on the plant ($E_o$), which observations have been erased ($E_-$), which observations have been inserted ($E_+$), and which observations have been enabled ($E_S$). The attacker observer $Obs_{att}(G)$ can be constructed using Algorithm~\ref{alg1}.

\begin{algorithm}[H]
\caption{Construction of the attacker observer $Obs_{att}(G)$}
\label{alg1}
\begin{algorithmic}[1]
\Require
Plant $G=(X,E,\Delta,x_0)$, $E_{ins}$, $E_{era}$, and $E_{ena}$.
\Ensure
Attacker observer $Obs_{att}(G)=(B,E_a,\Delta_{att},b_0)$.
\State Construct the observer $Obs(G)=(B,E_o,\Delta_o,b_0)$;
\State Let $E_a := E_o \cup E_+ \cup E_- \cup E_S$;
\State Let $\Delta_{att}:=\Delta_o$;
\ForAll{$e \in E_{ins}$,}
\ForAll{$b \in B$,}
\State $\Delta_{att}:=\Delta_{att} \cup (b,e_+,b)$;
\EndFor
\EndFor
\ForAll{$e \in E_{era}$,}
\ForAll{$b \in B$,}
\If{$\exists (b,e,b') \in \Delta_{att}$,}
\State $\Delta_{att}:=\Delta_{att} \cup (b,e_-,b')$;
\EndIf
\EndFor
\EndFor
\ForAll{$e \in E_{ena}$,}
\ForAll{$b \in B$,}
\If{$\exists (b,e,b') \in \Delta_{att}$,}
\State $\Delta_{att}:=\Delta_{att} \cup (b,e_S,b')$;
\EndIf
\EndFor
\EndFor
\end{algorithmic}
\end{algorithm}

Note that an analogous observer has been proposed in~\cite{Qi}. However, while in~\cite{Qi} we only consider sensors attack, here we modify the structure associated with both sensors and actuator enablement attacks.

We briefly explain how Algorithm~\ref{alg1} works. First, the observer $Obs(G)=(B,E_o,\Delta_o,b_0)$ is constructed. The set of states of  $Obs_{att}(G)$ coincides with the set of states of $Obs(G)$, as well as the initial state. The attack alphabet $E_a$ is defined. The transition relation is initialized at $\Delta_{att}:=\Delta_o$.  Indeed, all the events $e \in E_o$ are events that really happen in the plant. Therefore, when such events occur, $Obs_{att}(G)$ updates his states in accordance with the transition relation $\Delta_o$.

Then, for all events $e \in E_{ins}$ and for all states $b \in B$, we add self-loops $(b,e_+,b)$. Indeed, events $e_+ \in E_+$ are events that do not actually occur in the plant, thus $Obs_{att}(G)$ does not update his state.

Finally, for all events $e \in E_{era}$ and for all states $b \in B$, whenever a transition $(b,e,b')$ is defined, we add the transition $(b,e_-,b')$ to $\Delta_{att}$. Indeed, events $e_- \in E_-$ are events that occur in the plant even if they are erased by the attacker, thus $Obs_{att}(G)$ updates his state is the same way in case of $e$ and $e_-$. For all events $e \in E_{ena}$ and for all states $b \in B$, whenever a transition $(b,e,b')$ is defined, we add the transition $(b,e_S,b')$ to $\Delta_{att}$. Indeed, events $e_S \in E_S$ are events enabled by the attacker.

\begin{example} Consider the plant $G=(X,E,\Delta,x_0)$ and the observer in Fig.s~\ref{fig3}~(a) and (b), respectively. Let $E_o=E_c=\{a,b,c,g\}$, $E_{uo}=E_{uc}=\{d\}$, and $X_{us}=\{3\}$.

\begin{figure}[htbp]
  \centering
  \subfigure[$G$]{
    \includegraphics[width=1.5in]{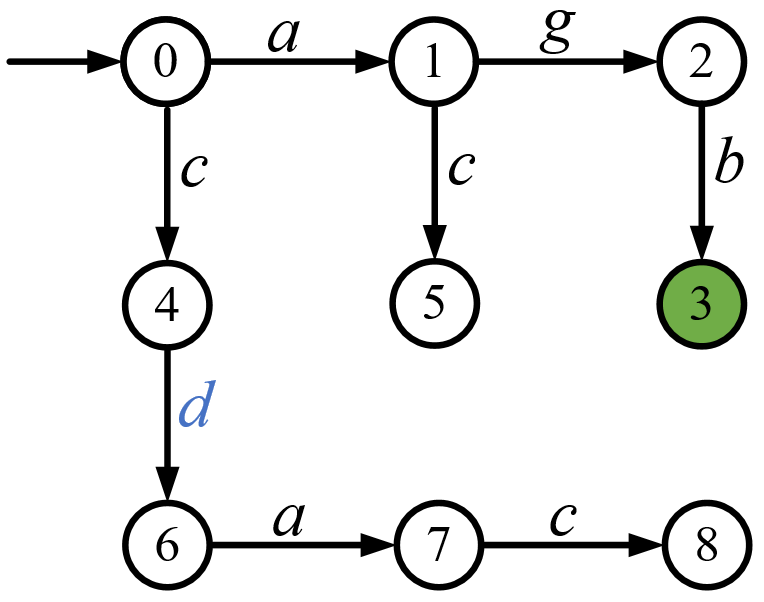}
  }
  \hspace{1in}
  \subfigure[$Obs(G)$]{
    \includegraphics[width=1.9in]{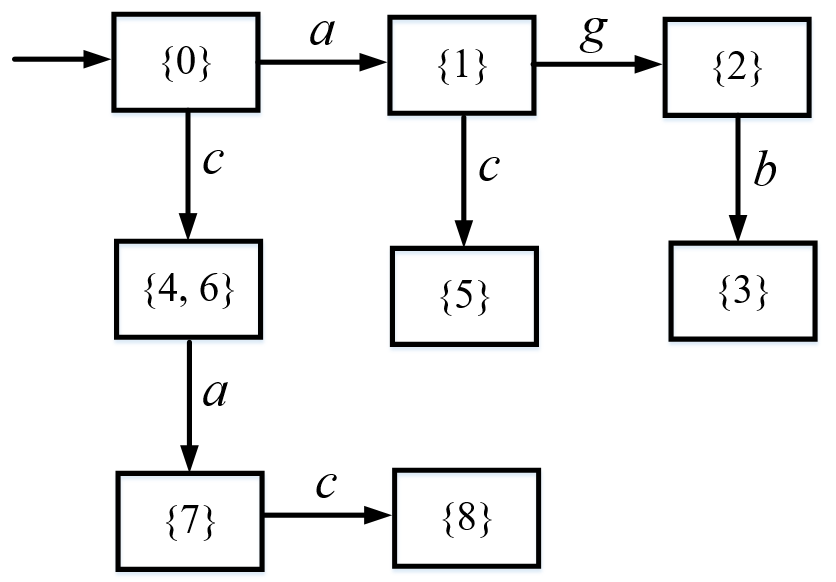}
  }

  \caption{(a) Plant $G$ and (b) Observer $Obs(G)$.}
  \label{fig3} 
\end{figure}

\begin{figure}[htbp]
  \centering
  \includegraphics[width=2.2in]{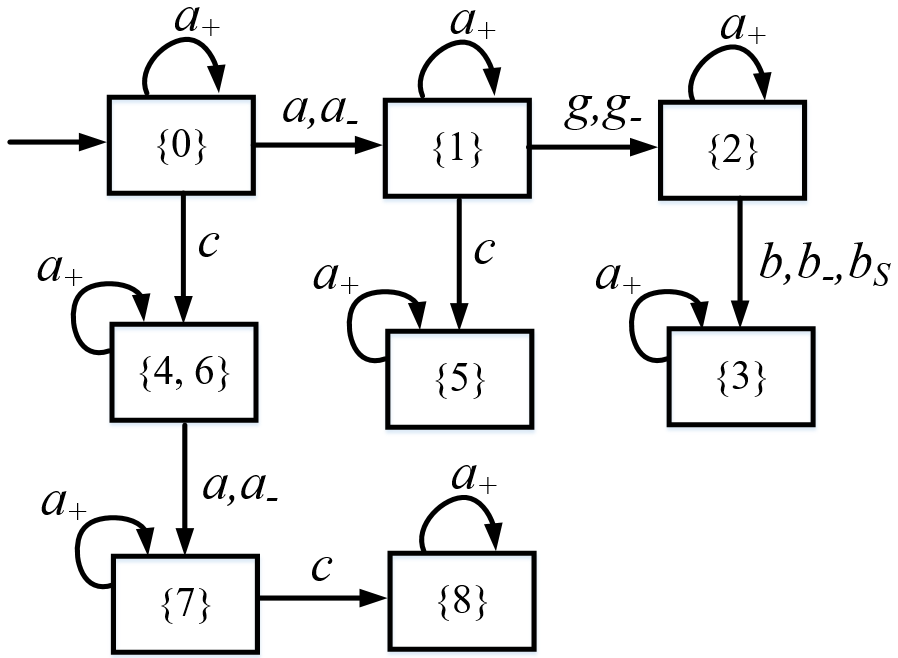}\\
  \caption{Attacker observer $Obs_{att}(G)$.}
  \label{fig5}
\end{figure}

Let $E_{ins}=\{a\}$, $E_{era}=\{a,b,g\}$, and $E_{ena}=\{b\}$. The attacker observer $Obs_{att}(G)$ is visualized in Fig.~\ref{fig5}.
Since $a \in E_{ins}$, we add self-loops labeled $a_+$ at all states. Since $a,b,g \in E_{era}$, arcs labeled $a$, $b$, and $g$, respectively, are also labeled  $a_-$, $b_-$, and $g_-$, respectively. Since $b \in E_{ena}$, the arc labeled $b$, is also labeled  $b_S$.~\hfill \qed
\label{exa1} \end{example}

\subsection{Supervisor Under Attack}\label{Supervisorunderattack}

The supervisor under attack $S_{P_a}$ provides the corrupted state estimation computed by the supervisor on the basis of the attack strings $w\in E_a^*$. The supervisor under attack $S_{P_a}$ generates two different sets of words. The first set contains all words on $E_a$ that may either result from an uncorrupted observation or from a corrupted observation which keeps the attacker stealthy. The second set of words contains all the previous words continued with a symbol in $E_a$ so that the resulting word is not consistent with an uncorrupted observation. While the words in the first set lead to a set of states that according to the supervisor are consistent with the perceived observation, those in the second set lead to a dummy state denoted as $y_\emptyset$. The supervisor under attack $S_{P_a}$ can be constructed using Algorithm~\ref{alg2}.

\begin{algorithm}[htbp]

\caption{Construction of the supervisor under attack $S_{P_a}$}
\label{alg2}
\begin{algorithmic}[1]
\Require
Plant $G=(X,E,\Delta,x_0)$, supervisor $S_P=(Y,E,\Delta_s,y_0)$, event sets $E_{ins}$, $E_{era}$, and $E_{ena}$.
\Ensure
Supervisor under attack $S_{P_a}=(Y_{a},E_a,\Delta_{sa},y_0)$.
\State Let $Y_a:=Y \cup \{y_{\emptyset}\}$;
\State Let $E_a := E_o \cup E_+ \cup E_- \cup E_S$;
\State Let $\Delta_{sa}:=\Delta_s$;
\State Construct the observer $Obs(G)=(B,E_o,\Delta_o,b_0)$;
\ForAll{$e \in E_{ena}$,}
\ForAll{$y \in Y$,}
\If{$\exists s \in E_o^*$: $(b_0,s,b) \in \Delta_o^*$, $(y_0,s,y) \in \Delta_s^*$, and $e \in \Gamma(b) \setminus \Gamma(y)$,}
\State $\Delta_{sa}:=\Delta_{sa} \cup (y,e_{S},y_\emptyset)$;
\If{$\exists s \in E_o^*$: $(b_0,s,b) \in \Delta_o^*$, $(y_0,s,y) \in \Delta_s^*$, $e \in \Gamma(b) \setminus \Gamma(y)$, and $e \in E_{ins}$,}
\State $\Delta_{sa}:=\Delta_{sa} \cup (y,e_+,y_{\emptyset})$;
\If{$\exists s \in E_o^*$: $(b_0,s,b) \in \Delta_o^*$, $(y_0,s,y) \in \Delta_s^*$, $e \in \Gamma(b) \setminus \Gamma(y)$, and $e \in E_{era}$,}
\State $\Delta_{sa}:=\Delta_{sa} \cup (y,e_-,y)$;
\EndIf
\EndIf
\EndIf
\EndFor
\EndFor
\ForAll{$e \in E_{o}$,}
\ForAll{$y \in Y$,}
\If{$\nexists (y,e,y') \in \Delta_{sa}$ $\wedge$ $\nexists s \in E_o^*$: $(b_0,s,b) \in \Delta_o^*$, $(y_0,s,y) \in \Delta_s^*$, and $e \in \Gamma(b) \setminus \Gamma(y)$,}
\State $\Delta_{sa}:=\Delta_{sa} \cup (y,e,y_\emptyset)$;
\EndIf
\EndFor
\EndFor
\ForAll{$e \in E_{ins}$,}
\ForAll{$y_a \in Y_a$,}
\If{$\exists (y_a,e,y_a') \in \Delta_{sa}$,}
\State $\Delta_{sa}:=\Delta_{sa} \cup (y_a,e_+,y_a')$;
\EndIf
\EndFor
\EndFor
\ForAll{$e \in E_{era}$,}
\ForAll{$y_a \in Y_a$,}
\If{$\exists (y_a,e,y_a') \in \Delta_{sa}$,}
\State $\Delta_{sa}:=\Delta_{sa} \cup (y_a,e_-,y_a)$;
\EndIf
\EndFor
\EndFor
\end{algorithmic}
\end{algorithm}

We briefly explain how Algorithm~\ref{alg2} works. Consider a supervisor $S_P=(Y,E,\Delta_s,$ $y_0)$. First, the set of states $Y_a$ is defined as $Y \cup \{y_{\emptyset}\}$, where $y_\emptyset$ is a dummy state: the supervisor detects the presence of an attacker when $y_\emptyset$ is reached. In addition, the attack alphabet $E_a$ is computed, and the transition relation of $S_{P_a}$ is initialized at $\Delta_s$.

At Step~4, the observer $Obs(G)=(B,E_o,\Delta_o,b_0)$ is constructed. Then, for all $e \in E_{ena}$, and for all $y \in Y$, if  the following condition holds:
\begin{itemize}
\item[(i)]
 there exists an observable sequence $s$ that leads to a state $b$ of the observer where $e$ is enabled, and to a state $y$ of the supervisor where $e$ is not enabled,
\end{itemize}
we add transition $(y,e_S,y_\emptyset)$ to $\Delta_{sa}$ (Step~8).

If $e$ satisfies the condition in the previous item, and it is also an event in $E_{ins}$ (resp., $E_{era}$), then we add transition $(y,e_+,y_{\emptyset})$ (resp., $(y,e_-,y)$) to  $\Delta_{sa}$ (Steps~10 and 12, respectively).

For all events $e \in E_o$, and for all states $y \in Y$, if the following two conditions hold:
\begin{itemize}
\item[(ii)]
a transition $(y,e,y')$ is not defined in $\Delta_{sa}$, and
\item[(iii)]
there does not exist an observable sequence $s$ that satisfies the condition in the item (i),
\end{itemize}
we add transition $(y,e,y_\emptyset)$ to $\Delta_{sa}$.

Finally, for all events $e \in E_{ins}$, and for all states $y_a \in Y_a$, whenever a transition $(y_a,e,y_a')$ is defined, we add  transition $(y_a,e_+,y_a')$ to $\Delta_{sa}$. Indeed, the supervisor cannot distinguish events $e_+$ from $e$. For all events $e \in E_{era}$, and for all states $y_a \in Y_a$, whenever a transition $(y_a,e,y_a')$ is defined, we add self-loop $(y_a,e_-,y_a)$. Indeed, the supervisor cannot observe $e_- \in E_-$.

\begin{example} Consider the observer and the supervisor in Fig.~\ref{fig3}~(b) and Fig.~\ref{fig6}~(a), respectively. Let $E_{ins}=\{a\}$, $E_{era}=\{a,b,g\}$, and $E_{ena}=\{b\}$. The supervisor under attack $S_{P_a}$ is visualized in Fig.~\ref{fig6}~(b).

\begin{figure}[htbp]
  \centering
  \subfigure[$S_P$]{
    \includegraphics[width=1.7in]{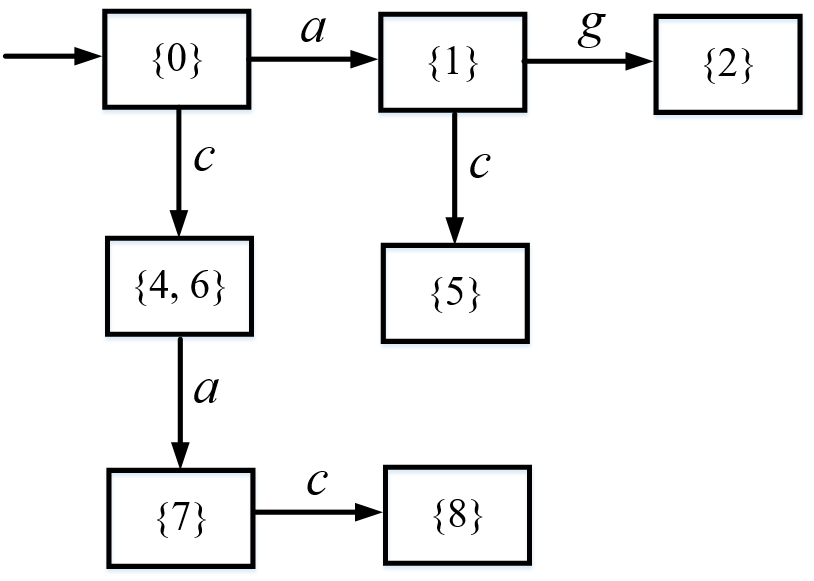}
  }
  \hspace{1in}
  \subfigure[$S_{P_a}$]{
    \includegraphics[width=1.7in]{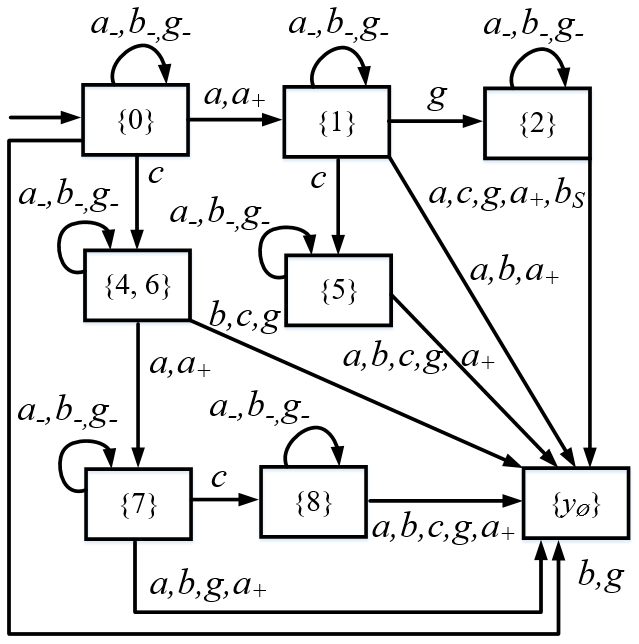}
  }

  \caption{(a) Supervisor $S_P$ and (b) Supervisor under attack $S_{P_a}$.}
  \label{fig6} 
\end{figure}

Since $b \in E_{ena}$, and $b$ is disabled by the supervisor at state $\{2\}$, we add a transition $(\{2\},b_S,\{y_\emptyset\})$. Since $b$ is also an event in $E_{era}$, then we add a self-loop $(\{2\},b_-,\{2\})$. Since $a,b,c,g \in E_o$, and transition labeled $a$ is not defined at state $\{1\}$, we add transition $(\{1\},a,\{y_\emptyset\})$. Similar arguments can be used to explain the other transitions labeled $a$, $b$, $c$, and $g$ that yield state $y_\emptyset$.

Since $a \in E_{ins}$, and there is a transition $(\{1\},a,\{y_\emptyset\})$ in $\Delta_{sa}$, then we add transition $(\{1\},a_+,\{y_\emptyset\})$ to $\Delta_{sa}$. Similar arguments can be used to explain the other transitions labeled $a_+$ in $\Delta_{sa}$.

Since $a,b,g \in E_{era}$, and there is transition $(\{1\},a,\{y_\emptyset\})$ in $\Delta_{sa}$, then we add a self-loop $(\{1\},a_-,\{1\})$ to $\Delta_{sa}$. Similar arguments can be used to explain the other self-loops labeled $a_-$, $b_-$ and $g_-$ in $\Delta_{sa}$.~\hfill \qed
\label{exa2} \end{example}

\section{ATTACK STRUCTURE}\label{Sectionattackstructure}

In this section, a particular structure called \emph{attack structure} is introduced. The notion of supremal stealthy attack substructure is also given.

\subsection{Definition}

The attack structure can be formally defined as follows.

\begin{definition} Consider the closed-loop system $S_P/G$ with set of compromised events $E_{com}$. Let $Obs_{att}(G)$ and $S_{P_a}$ be the attacker observer and supervisor under attack, respectively. The attack structure $A$ w.r.t. $S_P/G$ and $E_{com}$ is the DFA: $A=Obs_{att}(G) \| S_{P_a}$.~\hfill \qed
\label{def12} \end{definition}

\begin{definition} The set of target states of an attack structure $A=(R,E_a,\Delta_a,r_0)$ is $R_t:=\{r=(b,y_a) \in R \ | \ b \cap X_{us}\neq\emptyset\}$.~\hfill \qed \label{def21} \end{definition}

Target states are those states whose first entry contains an element in the set of unsafe states. The closed-loop system may reach the unsafe state when a target state is reached.

\begin{figure}[htbp]
  \centering
  \includegraphics[width=4in]{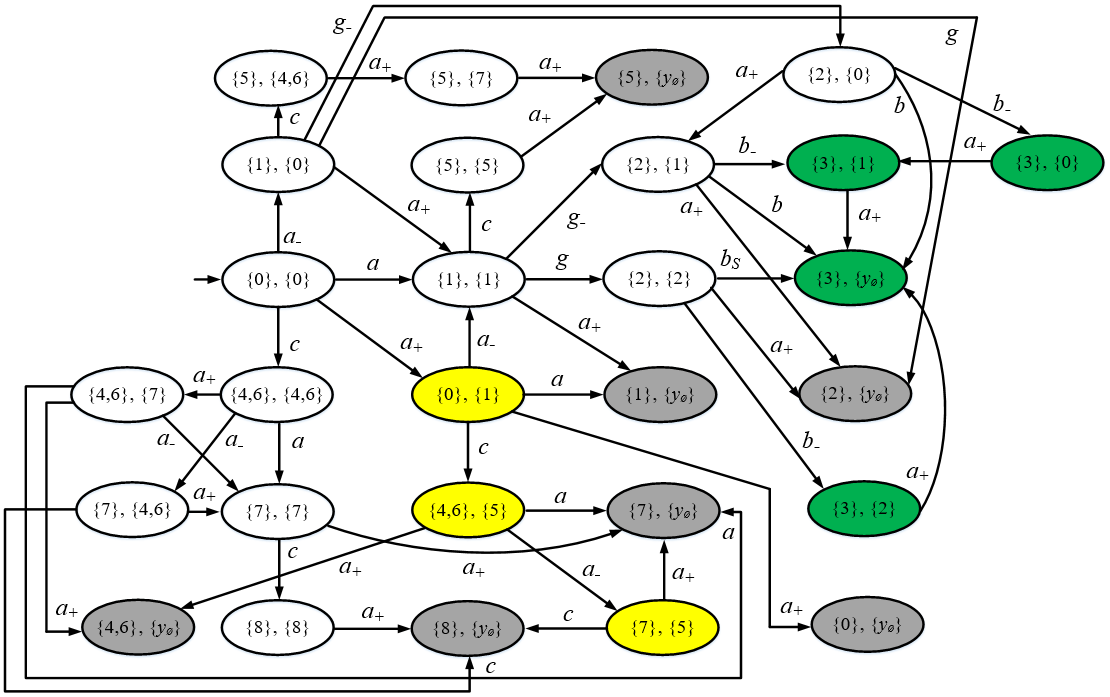}\\
  \caption{Attack structure $A$.}
  \label{fig7}
\end{figure}

\begin{example} Consider again the closed-loop system $S_P/G$ in Example~\ref{exa1} and Example~\ref{exa2}. $Obs_{att}(G)$ and $S_{P_a}$ are depicted in Fig.~\ref{fig5} and Fig.~\ref{fig6} (b), respectively. The attack structure $A=Obs_{att}(G) \| S_{P_a}$ is visualized in Fig.~\ref{fig7} (neglect the colours of the states for the time being). \label{exa3} \end{example}

At the initial state $(\{0\},\{0\})$, if event $a$ occurs in the plant, since $a \in E_{era}$, the attacker either erases $a$ or does not erase $a$, corresponding to transitions $[(\{0\},\{0\}),a_-,(\{1\},\{0\})]$ and $[(\{0\},\{0\}),a,(\{1\},\{1\})]$, respectively. Since $a \in E_{ins}$, the attacker may insert event $a$ at state $(\{0\},\{0\})$, this corresponds to transition $[(\{0\},\{0\}),a_+,(\{0\},\{1\})]$. Similar arguments can be used to explain the other states and transitions in $A$.

In $A$, four target states are highlighted in green: $(\{3\},\{0\})$, $(\{3\},\{1\})$, $(\{3\},$ $\{2\})$, and $(\{3\},\{y_\emptyset\})$, thus the attacker is effective. When the target state is reached, the plant is in unsafe state \{3\}.

In $A$, when state $(\{1\},\{1\})$ is reached, event $g$ may occur in the plant. If the attacker erases it, state $(\{2\},\{1\})$ is reached. Then $b$ may occur in the plant, corresponding to the transition $[(\{2\},\{1\}),b,(\{3\},\{y_\emptyset\})]$. Similar arguments can be used to explain how the other target states $(\{3\},\{0\})$ and $(\{3\},\{1\})$ are reached.

In $A$, when state $(\{2\},\{2\})$ is reached, the attacker can enable event $b$, corresponding to the transition $[(\{2\},\{2\}),b_S,(\{3\},\{y_\emptyset\})]$. If $b_S$ is erased, state $(\{3\},\{2\})$ is reached.~\hfill \qed

\subsection{Supremal Stealthy Attack Substructure}

We notice that a subset of states in $A$ reveals the presence of an attacker, thus the attacker should make sure these states are not reached if he wants to remain stealthy. We call \emph{stealthy attack substructure} the structure obtained by appropriately trimming $A$, and we show how to obtain a \emph{supremal stealthy attack substructure} that contains all the possible attacks that remain stealthy.

\begin{definition} The set of exposing states of an attack structure $A=(R,E_a,\Delta _a,$ $r_0)$ is $R_e:=\{r=(b, y_a)\in R \mid y_a=y_\emptyset\}$.~\hfill \qed  \label{def16} \end{definition}

Exposing states are those states whose second entry is equal to $y_\emptyset$. When such a state is reached, the corrupted observation is not consistent with the observation generated by the plant without attack, thus the supervisor detects that the system is under attack.

Further additional states should be removed from $A$, in particular, those from which an exposing state is surely reached after a finite number of events generated by the plant. We call the set of such states \emph{weakly exposing region} $\overline{R}_e$. The weakly exposing region can be computed using Algorithm~3 in \cite{Qi}.

\begin{example} Consider the attack structure $A$ in Fig.~\ref{fig7}, the exposing states are highlighted in gray, and the other states in $\overline{R}_e$ are highlighted in yellow. \label{exa4} \end{example}

As an example, since in the attack structure there exists transition $[(\{7\},\{5\}),$ $c, (\{8\},\{y_\emptyset\})]$ and $c \notin E_{era}$, and there does not exist a transition labeled $e_+ \in E_+$ yielding a state not in $\overline{R}_e$, then state $(\{7\}, \{5\})$ should be added to $\overline{R}_e$.~\hfill \qed

Let $A$ be an attack structure, and $\overline{R}_e$ be its weakly exposing region. The \emph{supremal stealthy attack substructure} $A^{ss}$ is obtained removing from $A$ all states in $\overline{R}_e$ and their input and output arcs (a more detailed discussion on this can be found in Section 7 of \cite{Qi}).

\begin{figure}[htbp]
  \centering
  \includegraphics[width=3.6in]{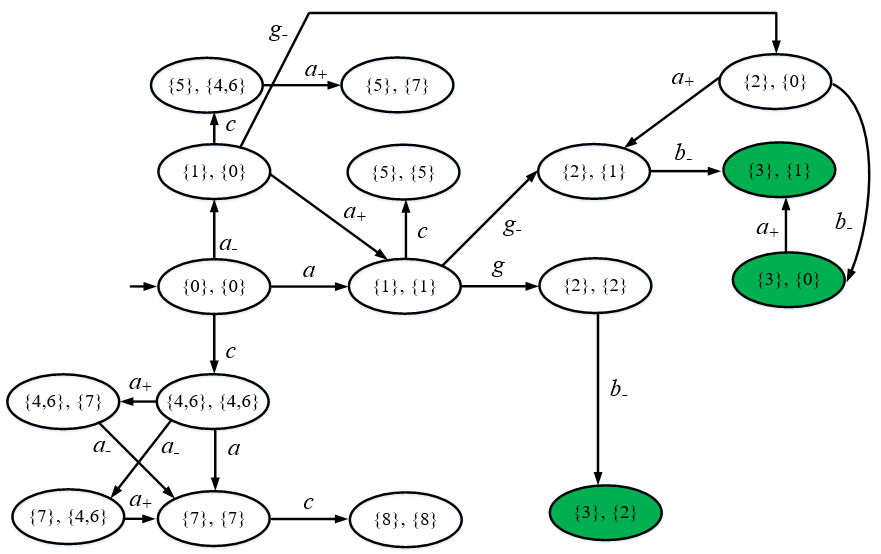}\\
  \caption{Supremal stealthy attack substructure $A^{ss}$.}
  \label{fig8}
\end{figure}

\begin{example} Consider again the attack structure $A$ in Fig.~\ref{fig7}. The supremal stealthy attack substructure $A^{ss}$ is visualized in Fig.~\ref{fig8}.
\label{exa5} \end{example}

At state $(\{2\},\{2\})$, if the attacker enables event $b$, as shown in $A$, the exposing state $(\{3\},\{y_\emptyset\})$ is reached, then the attack is not stealthy, thus state $(\{3\},\{y_\emptyset\})$ should be removed. If the attacker enables event $b$, and then erases it, as shown in $A^{ss}$, state $(\{3\},\{2\})$ is reached, thus the attack is stealthy.~\hfill \qed

\subsection{Computational Complexity} \label{subcomplexity}

Given a plant $G$ with set of states $X$. The observer can be constructed in $2^{|X|}$ steps, and the supervisor can be constructed in $2^{|X|}$ steps \cite{Lafortune:2008}. The attack structure is obtained by computing $A=Obs_{att}(G) \parallel S_{P_a}$. Therefore, $A$ can be constructed in $2^{|X|} \times 2^{|X|}$ steps. Since the supremal stealthy attack substructure $A^{ss}$ is obtained by appropriately refining $A$, thus the complexity of constructing $A^{ss}$ is $O(2^{2|X|})$.

\section{CONCLUSIONS AND FUTURE WORK}\label{Conclusions}
In this paper we considered the problem of cyber attacks in supervisory control systems. We developed a supremal stealthy attack substructure that allows us to select attacks that cause the closed-loop system $S_P/G$ to reach the unsafe state. The way the attacks are generated guarantees that the supervisor never realizes the presence of an attacker before the unsafe state is reached.

In the future, we will show how to synthesize a supervisor that is robust against such attacks.

%
%

\end{document}